\documentclass[twocolumn,showpacs,superscriptaddress,amsmath,amssymb]{revtex4}
\usepackage{stmaryrd}
\usepackage{mathrsfs}
\usepackage{graphicx}
\usepackage{dcolumn}
\usepackage{bm}

\def\bd{\begin{document}} \def\ed{\end{document}}
\def\bmp{\begin{minipage}} \def\emp{\end{minipage}}
\def\bcc{\begin{center}} \def\ecc{\end{center}}     \def\npg{\newpage}
\def\beq{\begin{equation}} \def\eeq{\end{equation}} \def\hph{\hphantom}
\def\be{\begin{equation}} \def\ee{\end{equation}} \def\r#1{$^{[#1]}$}
\def\n{\noindent} \def\ni{\noindent} \def\pa{\parindent}
\def\hs{\hskip} \def\vs{\vskip} \def\hf{\hfill} \def\ej{\vfill\eject}
\def\cl{\centerline} \def\ob{\obeylines}  \def\ls{\leftskip}
\def\underbar#1{$\setbox0=\hbox{#1} \dp0=1.5pt \mathsurround=0pt
   \underline{\box0}$}   \def\ub{\underbar}    \def\ul{\underline}
\def\f{\left} \def\g{\right} \def\e{{\rm e}} \def\o{\over} \def\d{{\rm d}}
\def\vf{\varphi} \def\pl{\partial} \def\cov{{\rm cov}} \def\ch{{\rm ch}}
\def\la{\langle} \def\ra{\rangle} \def\EE{e$^+$e$^-$} \def\pt{p_{\rm t}}
\def\dt{\delta}   \def\ie{{\it i.e.\;}}   \def\cf{{\it cf.\;}}
\def\bitz{\begin{itemize}} \def\eitz{\end{itemize}}
\def\btbl{\begin{tabular}} \def\etbl{\end{tabular}}
\def\btbb{\begin{tabbing}} \def\etbb{\end{tabbing}}
\def\beqar{\begin{eqnarray}} \def\eeqar{\end{eqnarray}}
\def\\{\hfill\break} \def\dit{\item{-}} \def\i{\item}
\def\bbb{\begin{thebibliography}{9} \itemsep=-1mm}
\def\ebb{\end{thebibliography}} \def\bb{\bibitem}
\def\bpic{\begin{picture}(260,240)} \def\epic{\end{picture}}
\def\akgt{\noindent{Acknowledgements}}
\def\fgn{\noindent{\bf\large\bf figure captions}}
\def\lan{\langle}
\def\ran{\rangle}
\def\p{\pi}
\def\ifmath#1{\relax\ifmmode #1\else $#1$\fi}%
\def\rc{\ifmath{{\mathrm{c}}}}
\def\cut{\ifmath{{\mathrm{cut}}}}
\def\rF{\ifmath{{\mathrm{F}}}}
\def\rK{\ifmath{{\mathrm{K}}}}
\def\rp{\ifmath{{\mathrm{p}}}}
\def\rt{\ifmath{{\mathrm{t}}}}
\def\LAB{\ifmath{{\mathrm{LAB}}}}
\def\cut{\ifmath{{\mathrm{cut}}}}
\def\beq{\begin{equation}}
\def\eeq{\end{equation}}
\def\us{^{(s)}}  \def\bea{\begin{eqnarray}} \def\eea{\end{eqnarray}}
\def\nbr{\nonumber} \def\e{{\rm e}} \def\dt{\delta} \def\D{\Delta}
\def\r{\rho}  \def\unln{\underline}
\newcommand{\cinst}[2]{$^{\mathrm{#1}}$~#2\par}
\newcommand{\crefi}[1]{$^{\mathrm{#1}}$}
\newcommand{\crefii}[2]{$^{\mathrm{#1,#2}}$}
\newcommand{\crefiii}[3]{$^{\mathrm{#1,#2,#3}}$}
\newcommand{\HRule}{\rule{0.5\linewidth}{0.5mm}}
\def\yw{Y_{\rm w}}  \def\ew{\eta_{\rm w}}  \def\de{\delta\eta} \def\Bs{B_{\rm s}}
\def\bs{\boldsymbol}  \def\lm{\lambda}
\bd

\title{Examining the crossover from hadronic to partonic phase in QCD}

\author{Xu Mingmei, Yu Meiling , Liu Lianshou}
\email{liuls@iopp.ccnu.edu.cn}

\affiliation{Institute of Particle Physics, Huazhong Normal
University, Wuhan 430079, China}

\def\ssnn{\sqrt {s_{NN}}}

\begin{abstract}
It is argued that, due to the existence of  two vacua ---
perturbative and physical --- in QCD, the mechanism for the
crossover from hadronic to partonic phase is hard to construct. The
challenge is: how to realize the transition between the two vacua
during the gradual crossover of the two phases. A possible solution
of this problem is proposed and a mechanism for crossover,
consistent with the principle of QCD, is constructed. The essence of
this mechanism is the appearance and growing up of a kind of
grape-shape perturbative vacuum inside the physical one. A dynamical
percolation model based on a simple dynamics for the delocalization
of partons is constructed to exhibit this mechanism. The crossover
from hadronic matter to sQGP as well as the transition from sQGP to
wQGP in the increasing of temperature is successfully described by
using this model with a temperature dependent parameter.
\end{abstract}


\maketitle


\def\tc{T_{\rm cr}}
The theory of strong interaction ------ quantum chromodynamics QCD
has a complicated phase structure\;\cite{phdiagram}. It has been
shown by lattice gauge theory\;\cite{lattice} that at low
temperature and density there is a confined, chiral symmetry
breaking phase with hadrons as basic elements, while at high
temperature and density is the deconfined, symmetry restoration
phase with partons (quarks and gluons) as basic degree of freedom.
It is found that at zero baryon density and high enough temperature,
the transition from  hadronic to partonic matter is of a cross-over
type\;\cite{crossover}.

Crossover is a gradual change of the system from one phase to the
other without a definite transition point. The crossover between
hadronic and partonic phases at high temperature has been firmly
settled by lattice-QCD from thermodynamic argument, but really what
happens in crossover; how does the system crossover from one phase
to the other, are still open questions.

In this respect, it is worthwhile to consider a similar problem in
QED --- the Mott transition in atomic gas, where a neutral atom gas
is transformed to electro-magnetic plasma through the ionization of
atoms one by one. However, such a mechanism is inapplicable to QCD,
because the crossover of  a hadron matter to quark-gluon plasma
through the decomposition of hadrons to quarks (anti-quarks) one by
one contradicts {\it color confinement}. According to color
confinement, isolated color object cannot exist in physical vacuum.
The energy of an isolated color object in physical vacuum is
infinite\;\cite{QCDtext}.

Unfortunately, most of the current models on the market, which
claimed crossover\;\cite{claim}\cite{qmd}, have not taken this color
confinement property of QCD into proper account. In these models the
initial parton system evolutes and hadronizes gradually. The
hadronization condition is either that a parton ceases to interact
with other partons\;\cite{ampt} or that the total color interaction
from a pair (or a three particle state) of quarks with the remaining
system vanishes\;\cite{qmd}. In both cases the hadronization is
carried out one by one. After a long time, the system in
consideration is dominated by hadrons but there are still a few
partons moving in the physical vacuum of hadronic
matter\;\cite{YuLiuPRC}, violating the confinement property of QCD.

The difficulty of crossover in QCD lies in the fact that QCD has
different vacua --- perturbative and physical --- with partons and
hadrons as basic element, respectively. How to realize the
transition from one vacuum to the other during the gradual change
--- crossover --- between partonic and hadronic phases is a big
challenge\;\cite{biro}.
In this letter we will discuss this problem and try to find a
possible mechanism for the crossover in QCD compatible with the
principle of color confinement.

For this purpose let us turn to another kind of model
--- the geometrical bond percolation model\;\cite{kim}. This
model has no dynamical prescription but is nevertheless enlightening
in constructing a crossover model for QCD\;\cite{baym, satz}.

In the bond-percolation model, crossover is realized through the
{\it formation of clusters}. In the lattice-version of the model,
``a cluster is defined as a group of nearest-neighboring occupied
sites that are linked by occupied bonds. The {\it cluster size} is
the {\it number of sites} in a cluster''\;\cite{kim}. In our case,
the sites are distributed continuously in the space and will be in
the following referred to as {\it cells}. The cluster-formation
could be understood as: originally, isolated cells are color singlet
hadrons, when they are connected by bonds to form clusters, color
can flow among them through bonds, and only the cluster as a whole
keeps to be color singlet.

Clusters could be of various sizes. The formation of an {\it
infinite cluster}, which is defined in a finite system as a cluster
extending from one boundary to the other, \ie a cluster having, at
least in one dimension, the system size, is taken as the appearance
of a new constituent in the system. At that time there are still a
lot of clusters of various-sizes, cf. Fig.\;1(a), \ie the system as
a whole has not yet turned to the new phase. So we consider the
appearance of an infinite cluster as the {\it starting point of the
crossover} of the system to a new phase. The {\it crossover process
is completed when all the cells in the system are connected to a
unique cluster}, \cf Fig.\;1(b). In this process no isolated color
object will appear in physical vacuum, and thus no contradiction
with QCD.

\begin{figure}
\includegraphics[width=3.4in]{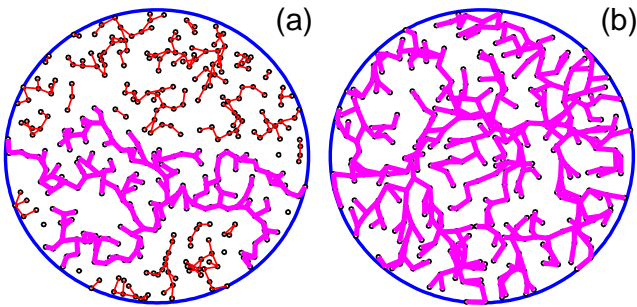}
\caption{\label{Fig. 2}  (Color online) Continuously distributed
cells connected by bonds form clusters. In (a) the big cluster
extending from the left- to the right-boundary is an {\it infinite
cluster}. In (b) all the cells are connected to an infinite
cluster.}
\end{figure}

Inspired by this achievement we propose our basic assumption as the
following. {\it The aggregation of hadrons could be of two forms:
gas-like or molecular-like}, \cf Fig.\;2\;\cite{celik}. In the first
case all the hadrons in an aggregation are melted together, forming
a big ``bag'' with quarks and gluons moving inside, which is usually
referred to as {\it QGP droplet}. In the second case, the hadrons
(or cells) in a cluster, like the atoms in a molecular, in spite of
being no longer color singlets, still keep their individuality, \ie
still occupy separate space regions, connected with each other by
bonds, forming a color singlet cluster.

\begin{figure}
\includegraphics[width=3.4in]{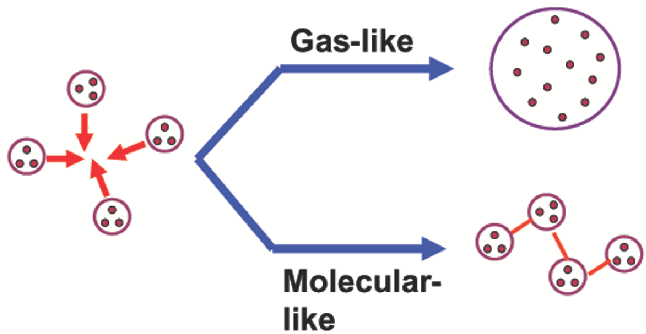}
\caption{\label{Fig. 3} (Color online) Two possible forms of nucleon
(or hadron) aggregation: gas-like and molecular-like. In both cases
only the whole aggregation is color singlet.}
\end{figure}

The cells, \ie the separated space regions occupied by the colored
parton-groups, are essentially ``holes'' dug by the groups of
colored parton in physical vacuum\;\cite{QCDtext}. These holes can
be viewed as potential wells if we add energy as an extra ordinate.
Partons can tunnel through the potential barriers between
neighboring wells all over the color-singlet cluster, which is
referred to as {\it delocalization}\;\cite{wangfan}.

When the crossover process is completed, all the wells in the system
are connected to a unique cluster, forming a {\it grape shape}
perturbative vacuum\;\cite{grape} of the nuclear size in 3-D, \cf
Fig.1(b). Meanwhile, there are still isolated bubbles of physical
vacuum in the system\;\cite{satz}, \cf the white regions in
Fig.1(b). The groups of colored parton are confined in the separate
wells, or cells, inside the {\it grape shape} perturbative vacuum,
being able to exchange color among themselves via quantum tunneling
through the potential barriers.

It could be expected that, the physical-vacuum bubbles can not
survive long in the environment of the grape-shape perturbative
vacuum and will soon be transformed to  perturbative too. The groups
of colored parton can then move around in the whole system,
resulting in a quark gluon matter possessing the property of {\it
perfect fluid}, which is conventionally referred to as {\it strongly
coupled QGP}, sQGP\;\cite{sqgp-exp, sqgp-th}.

At still higher temperature the potential barriers between
neighboring wells will drop to zero and all the wells disappear. In
this case all the parton-groups will be disassembled to independent
partons and the sQGP turned to {\it weakly coupled QGP}, wQGP.

Let us try to construct a simple model for realizing the above
assumptions and arguments. Our aim is to answer the question: {\it
how to crossover in QCD}. For simplicity, we take the initial system
to be a nucleon gas.

The model Hamiltonian for the 6 quarks in two near-by cells is
assumed to be\;\cite{wangfan} \beq
\mathcal{H}=\sum_{i=1}^{6}\f(m_{i}+\frac{p_{i}^{2}}{2m_{i}}\g)-T_{\rm
cm}+\sum_{i<j}V_{ij}^{C}, \eeq where $T_{\rm cm}$ is the
center-of-mass kinetic energy. $V_ {ij}^{C}$ is the color
interaction, which will be chosen as a square-confinement potential
$V_{ij}^{C}=-a_{c}\vec{\lambda}_{i}\cdot\vec{\lambda}_{j}r_{ij}^{2}$
when the quarks $i,j$ belong to one and the same cell. When they
belong to two nearby cells, the infinite potential in between will
drop down, forming a potential barrier, and a parametrization
$V_{ij}^{C}
=-a_{c}\vec{\lambda}_{i}\cdot\vec{\lambda}_{j}\frac{1-e^{-\mu
r_{ij}^{2}}}{\mu}$ will be used, where $\mu$ is a model parameter.
It turns out that as the increasing of $\mu$ the maximum distance
$S_0$ for delocalization increases, cf. Fig.\;3 below. Since at
higher temperature the quarks will be more free to move, i.e. $S_0$
will be larger, we require $\mu$ to be an ascending function of $T$.
From dimensional consideration we assume $\mu\propto T^2$.

In doing variational calculation the trial wave function of the
two-cell system  in adiabatic approximation is chosen to be an
antisymmetric six-quark product state\;\cite{wangfan}
\begin{equation}
    \f|\Psi_6(S)\g>=\mathcal {A}\f[\prod_{i=1}^{3}\psi_{\rm L}(\bs{r}_{i})
    \prod_{i=4}^{6}\psi_{\rm R}(\bs{r}_{i})\eta_{I_1 S_1}^{B_1}\eta_{I_2
    S_2}^{B_2}\chi_{c}^{B1}\chi_{c}^{B_2}\g]_{00},
\end{equation}
where $\mathcal {A}$ is the anti-symmetrization operator, which
permutes quarks between the two cells; $[\cdots]_{00}$ means that
the spin, isospin and color of the two cells are coupled to a
particular color singlet state with total spin and isospin equal
zero. For the orbital motion, we have the left (right) single-quark
orbital wave function $ \phi_{L}(\bs{r}_{i})=\f(\frac{1}{\pi
b^2}\g)^{\frac{3}{4}}\e^{-\frac{(\bs{r}_{i}+\frac{\bs{S}}{2})^2}{2b^2}},\
\phi_{R}(\bs{r}_{i})=\f(\frac{1}{\pi
b^2}\g)^{\frac{3}{4}}\e^{-\frac{(\bs{r}_{i}-\frac{\bs{S}}{2})^2}{2b^2}},$
where $\pm\frac{\bs{S}}{2}$ are the cell-centers, $S$ is the
distance between the two cells. $b$ is a baryon-size parameter.
Delocalized orbit is defined as
$\psi_{L}(\bs{r})=\frac{1}{N}[\phi_{L}+\epsilon\phi_{R}],\
  \psi_{R}(\bs{r})=\frac{1}{N}[\epsilon\phi_{L}+\phi_{R}],$
where $\epsilon$ is a variational parameter characterizing the
degree of delocalization and $N$ a normalization factor. At each
separation $S$, $\epsilon$ is determined by minimizing the energy
$E(S)=\frac{\f<\Psi_6(S)\f|\mathcal
{H}\g|\Psi_6(S)\g>}{\f<\Psi_6(S)|\Psi_6(S)\g>}.$ The model
parameters are: $m_{\rm u}=m_{\rm d}=313~{\rm MeV}$, $b=0.603~{\rm
fm}$, $a_{c}=101.14~{\rm MeV}/{\rm fm}^2$, while $\mu$ is leaving as
a free parameter.

It turns out that the model gives a reasonable amount of
delocalization. When the two cells are close together, \ie $S$ is
small, the delocalization is large $\epsilon$=1, but when the two
cells separate and $S$ increases to a certain distance $S_0$, the
delocalization degree $\epsilon$ suddenly drops to zero. The
dependence of $S_{0}$ on parameter $\mu$ is shown in
Fig.~\ref{fig:Fig3}.

\begin{figure}[!ht]
\includegraphics[width=3.6in]{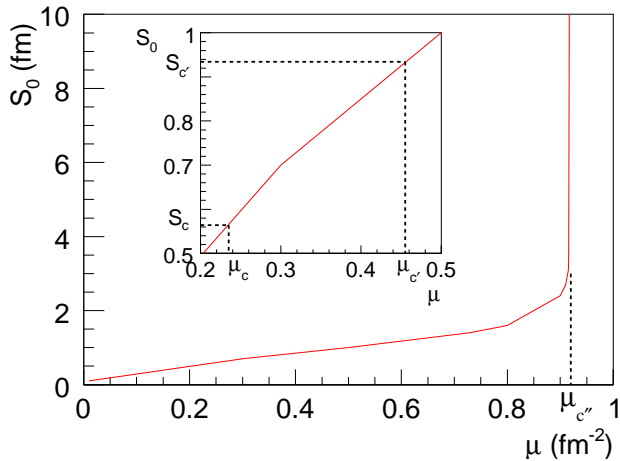}
\caption{The $\mu$ dependence of $S_0$ determined by the dynamics of
our model. Small pad shows a local part which is useful in
determining the critical values of parameter $\mu$.}
\label{fig:Fig3}
\end{figure}

$S_0$ provides us a distance within which the quarks have equal
probability to be simultaneously in the two cells, \ie within which
bond is likely to be formed. We will use it to construct a
bond-percolation model to realize the crossover in QCD. For
simplicity, we will consider the system to be a 2 dimensional static
one.

The initial configuration consists of $2\times197$ cells, which are
small spheres (circles in 2-D) of hard-core radius $r_{e}=0.1$ fm
distributed randomly in a big sphere (circle in 2-D) of radius $R=7$
fm. A cell with center departing from the center of the big sphere
(circle) farther than $R-r_{e}$, is considered as a {\it boundary
cell}.

The percolation procedure is as follows:

  1) Randomly select a cell $\alpha$ as a {\it mother cell}.

  2) Define the cells with $|\bs r-\bs r_\alpha|\leq S_0$ as
  {\it bond-candidate cells}, randomly select 3 of them
to form bonds connected to the mother cell $\alpha$, and call the
latter as {\it daughters}. If the number of candidate-cells is less
than 3, then the number of daughters is equal to the candidate
number.


  3) For every daughter of cell $\alpha$
  randomly select 2 bond-candidate cells of it to form
  bonds, and call them granddaughters.
  Repeat the procedure until no bond-candidate cell can be found anymore.


   4) Then choose another cell $\beta$ from the left
unbounded cells as another mother cell, and repeat the procedure
starting from step 2.

   5) Repeat step 4 until all the cells are exhausted.

In this way, every cell is assigned to a cluster. In each cluster,
find the boundary cells if any, and calculate the distance between
every two boundary cells. Denote the maximum distance by $d$. A
cluster with $d>\sqrt{2}R$ is taken as an {\it infinite cluster}.

\begin{figure}[!ht]
\includegraphics[width=3.5in]{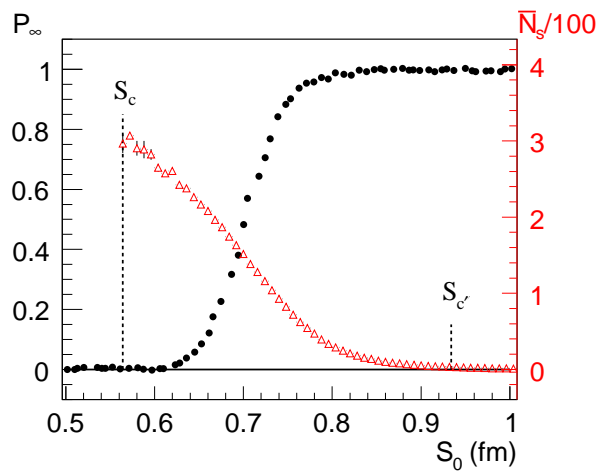}
\caption{ (Color online) Numerical results $P_{\infty}$ (full
circles) and $\bar N_s/100$ (open triangles) versus $S_0$ for 10 000
events. The dashed lines correspond to the 2 thresholds, i.e. $T_c$
the place where $P_\infty$ starts to increase from zero, $T_{c'}$
where the mean number $\bar N_s$ of cells outside an infinite
cluster tends to zero. }
\end{figure}

A configuration with all the cells assigned to clusters is called an
{\it event}. Generate $N$ events, and determine in each event
whether there are infinite cluster(s). Suppose $M$ is the number of
events with infinite cluster(s), the {\it probability for the
appearance of event with infinite-cluster(s)} can be expressed as
$P_\infty=\lim_{N\to\infty} {M}/{N}$. The dependence of $P_\infty$
on $S_0$ is plotted in Fig.\;4 as full circles. We see that at a
certain value $S_0=S_c$, $P_\infty$ starts to increase from zero,
and the system starts to crossover to a new phase. The corresponding
value of $\mu$ is $\mu=\mu_c$.

Let $N_s$ be the number of cells outside of an infinite cluster in
an event. The mean $\bar{N_s}$ versus $S_0$ is plotted in Fig.\;4 as
open triangles. The point where $\bar{N_s}=0$ marks the
accomplishment of crossover and will be denoted by $S_0=S_{c'}$. The
corresponding value of $\mu$ is $\mu_{c'}$, \cf the dashed lines in
Fig.\;4 and in the small pad of Fig.\;3.

The starting and ending points $\mu_c$ and $\mu_{c'}$ of crossover
can be extracted from Fig's.\;3, 4. Assuming $\mu\propto T^2$ we get
the corresponding temperatures: $T_{c'}\approx 1.39\, T_c$.

It is interesting to see from Fig.\;3 that as $\mu$ increases
further from $\mu_{c'}$, at a certain point $\mu_{c''}$,
$S_0$ increases sharply to infinity. An infinite $S_0$ means that
partons can flow from one well to the other no matter how far away
they are. This means that at this point the potential barriers have
dropped to zero, and there is no well in the vacuum anymore. Partons
can move freely inside the whole system, and the sQGP turns to wQGP.
Extracting the point $\mu_{c''}$ for this transition from the figure
and assuming again $\mu\propto T^2$, we get $T_{c''}\approx 1.98
T_c$\;\cite{tcpp}.

In this letter a possible mechanism for the crossover in QCD,
compatible with the principle of color confinement, is proposed. The
way for vacuum-change during the gradual crossover between the two
phases is via the following steps.

1)\ As the increasing of temperature, partons in hadrons start to be
delocalized, tunneling between neighboring hadrons, and change the
latter to {\it colored parton-groups} located in potential wells.

2)\ The wells connected by tunnels form color singlet clusters,
which grow up in the physical vacuum, and eventually become a big
cluster of the system size. Inside this cluster is a grape-shape
perturbative vacuum, while outside of it is physical vacuum bubbles,
which will change to perturbative as the further increase of
temperature, and the colored parton-groups in the wells can then
move around, forming a fluid-like matter --- sQGP.

3)\ The potential barriers will drop to zero at a higher temperature
and all the wells disappear, the sQGP is then turned to wQGP.

Last but not least, a dynamical percolation model is constructed for
the first time, which is superior to the purely geometrical
percolation models. 1)\  Our model has dynamical foundation and the
cells and bonds have definite physical meaning. 2)\  We use a
maximum length $S_0$ for bond-formation, obtained from dynamical
calculation, and construct 3 bonds {\it randomly} within this
length. This is to substitute the bond-formation probability $p$ put
in the geometrical models by hand. 3)\ Our model has a
temperature-dependent parameter $\mu$ and is able to deal with the
crossover as well as the transition from sQGP to wQGP in the
increasing of temperature, while the geometrical percolation models
are able to treat only the effect of the increasing of density.

In the proposed scenario the crossover is accomplished within a
temperature range from $T_c$ to $T_{c'}$. It could be expected that
at some energies lower than 200 GeV, the produced dense matter may
be at a mediate stage of crossover, which in our scenario consists
of an infinite cluster accompanied by a large number of other
clusters, cf. Fig.\;1(a). To study the properties of such a
structure of matter theoretically and experimentally is worthwhile,
which will deepen our understanding on QCD phase diagram and
critical dynamics.


We thank Wang Fan, Ping J. L., Li J. R., Wu Y. F., Zhuang P. F., H.
C. Ren, J. Shock, N. Xu and H. Z. Huang for helpful discussions.
This work is supported by NSFC under projects-10775056, 90503001.

\ed
\begin{thebibliography}{99}
\bibitem{grape} There is non-perturbative barrier-tunelling
in the grape-shape vacuum, but except that it is perturbative.
\end{thebibliography}
